\documentclass[12pt]{article}

\begin{document}

\title{\normalsize \hfill UWThPh-2000-31 \\[1cm] \LARGE
Seesaw model with softly broken $L_e - L_\mu - L_\tau$}

\author{L.\ Lavoura \\
\small Universidade T\'ecnica de Lisboa \\
\small Centro de F\'\i sica das Interac\c c\~oes Fundamentais \\
\small Instituto Superior T\'ecnico, P--1049-001 Lisboa, Portugal
\\[3mm]
W.\ Grimus \\
\small Universit\"at Wien, Institut f\"ur Theoretische Physik \\
\small Boltzmanngasse 5, A--1090 Wien, Austria}

\date{2 August 2000}

\maketitle

\begin{abstract}
We enlarge the lepton sector of the Standard Model
by adding to it two right-handed neutrino singlets, and
propose a model based on the seesaw mechanism
and on the lepton number $\bar L = L_e - L_\mu - L_\tau$.
The model is innovative from the theoretical point of view in that the
$\bar L$-invariant mass term
of the right-handed neutrinos
is associated with the large mass scale responsible for the seesaw mechanism,
whereas $\bar L$
is softly broken at a smaller scale by
the remaining Majorana mass terms of the right-handed neutrinos. 
While being very economical,
the model has predictive power:
one neutrino is massless ($m_3 = 0$),
the lepton mixing matrix $U$ features $U_{e3} = 0$,
and the solar-neutrino oscillations have maximal amplitude.

\end{abstract}

\vspace{6mm}

The recent results of Super-Kamiokande \cite{SK-atm},
providing evidence for a dist\-ance-depen\-dent
depletion of the muon neutrinos produced in the atmosphere by cosmic rays,
have provided strong evidence for neutrino oscillations and thus for massive
neutrinos \cite{review}. 
In this context,
one would like to understand why
are neutrinos so much lighter than the charged leptons;
a possible solution to this puzzle
is given by the seesaw mechanism \cite{seesaw}.
One would also like to be able to derive the characteristic features
of the neutrino mass spectrum and of the lepton mixing matrix 
from a model \cite{barr}.
In particular,
one knows that the neutrino mass-squared differences 
$\Delta m^2_{\rm atm}$ and $\Delta m^2_\odot$ corresponding to
atmospheric and solar-neutrino oscillations,
respectively, are of different orders of magnitude:
\begin{equation} \label{hierarchy}
3 \times 10^{-3}\, {\rm eV}^2 \sim
\Delta m^2_{\rm atm} = \left| m_3^2 - m_2^2 \right|
\gg \Delta m^2_\odot = \left| m_2^2 - m_1^2 \right|
\sim 10^{-7}\, {\rm eV^2}.
\end{equation}
For definiteness,
we have used for $\Delta m^2_\odot$ a value typical
of one of the currently experimentally favored solutions
of the solar-neutrino problem \cite{vagins},
namely the LOW solution \cite{LOW}.
Furthermore, disregarding the possibility of a nearly degenerate
neutrino mass spectrum, 
there are two spectral types:
either $m_3 \gg m_2 \gg m_1$ (conventional hierarchy)
or $m_3 \ll m_2 \approx m_1$ (inverted hierarchy).
At present, experiment cannot distinguish between these two cases.
As for the lepton mixing matrix $U$,
one of its prominent features is the smallness of $U_{e3}$ \cite{CHOOZ, Ue3},
for which one would like to find an explanation.
One also knows that
$\left| U_{\mu 3} \right| \approx \left| U_{\tau 3} \right|$
(maximal mixing of the atmospheric neutrinos) and that,
in the most viable solutions of the solar-neutrino problem,
$\left| U_{e1} \right| \approx \left| U_{e2} \right|$
(large mixing of the solar neutrinos).

It has been pointed out \cite{barbieri}
that the assumption of an approximate lepton-number symmetry
$\bar L \equiv L_e - L_\mu - L_\tau$
would constitute a good starting point
for a model of the lepton mass matrices.
Indeed,
if there are no $\left| \Delta \bar L \right| = 2$ mass terms,
then one predicts an inverted hierarchy of the neutrino masses,
with $m_3 = 0$ and $m_2 = m_1$;
moreover,
$\left| U_{e1} \right| = \left| U_{e2} \right| = 1 / \sqrt{2}$
and $U_{e3} = 0$.
On the other hand,
$\bar L$ must be slightly broken,
since we need $m_1$ to be different from $m_2$
in order to have a non-zero $\Delta m^2_\odot$.
For models of this type see Ref.~\cite{models}.

In this letter we propose a model which combines the seesaw mechanism
and a softly broken U(1) symmetry, corresponding to
$\bar L$, in a minimal fashion.
Our model,
while being extremely simple and economical,
retains all the essential predictions of the $\bar L$-symmetric situation
reviewed in the previous paragraph.
It is given by the Standard Model of the electroweak interactions,
based on the gauge group SU(2)$\times$U(1)
and with a single Higgs doublet 
$\phi = \left( \varphi^+, \varphi^0 \right)^T$.
The vacuum expectation value (VEV) of $\varphi^0$ is $v / \sqrt{2}$.
We also introduce two right-handed neutrino singlets,
$\nu_{R1}$ and $\nu_{R2}$.
We have the following assignments of $\bar L$ to these multiplets:
\begin{equation} \label{Lbar}
\renewcommand{\arraystretch}{1.4}
\begin{array}{c|cccccc}
& \nu_e, e & \nu_\mu, \mu & \nu_\tau, \tau & \nu_{R1} & \nu_{R2} &
\phi \\ \hline
\bar L & 1 & -1 & -1 & 1 & -1 & 0 \end{array}\ .
\end{equation}
The Yukawa Lagrangian is, therefore, given by
\begin{eqnarray}
{\cal L}_{\rm Y} &=& - \frac{\sqrt{2}}{v}
\left( \begin{array}{cc} \varphi^0, & - \varphi^+ \end{array} \right)
\left[ a\, \bar\nu_{R1}
\left( \begin{array}{c} \nu_{eL} \\ e_L \end{array} \right)
+ b^\prime \bar\nu_{R2}
\left( \begin{array}{c} \nu_{\mu L} \\ \mu_L \end{array} \right)
+ b^{\prime \prime} \bar\nu_{R2}
\left( \begin{array}{c} \nu_{\tau L} \\ \tau_L \end{array} \right)
\right]
\nonumber\\[1mm]
 & &
- \frac{\sqrt{2}}{v}
\sum_{\ell = e, \mu, \tau} m_\ell
\left( \begin{array}{cc} \bar\nu_{\ell L}, & \bar\ell_L \end{array} \right)
\ell_R 
\left( \begin{array}{c} \varphi^+ \\ \varphi^0 \end{array} \right) 
+ {\rm H.c.},
\label{yukawa}
\end{eqnarray}
where $a$,
$b^\prime$,
and $b^{\prime \prime}$ are complex coupling constants,
while the $m_\ell$ ($\ell = e, \mu, \tau$)
are the charged-lepton masses.
The first line of Eq.~(\ref{yukawa}) shows the Yukawa couplings
of the right-handed neutrino singlets.
The second line displays the Yukawa couplings
which give mass to the charged leptons.
Notice that we have taken,
without loss of generality,
the latter to be flavor-diagonal.

Since the right-handed neutrino fields are gauge singlets,
we can write down the mass term 
\begin{equation} \label{LM}
{\cal L}_M = - M \bar\nu_{R1} C \bar\nu_{R2}^T + {\rm H.c.},
\end{equation}
which is compatible with the lepton number $\bar L$.
In Eq.~(\ref{LM}),
$C$ is the (antisymmetric) charge-conjugation matrix.
The mass $M$ will play the role of a large seesaw scale.

We now assume $\bar L$ to be softly broken
by the following $\left| \Delta \bar L \right| = 2$ Majorana mass terms:
\begin{equation} \label{soft breaking}
{\cal L}_{\rm sb} = - {\textstyle \frac{1}{2}} \left(
R \bar\nu_{R1} C \bar\nu_{R1}^T +
S \bar\nu_{R2} C \bar\nu_{R2}^T \right) + {\rm H.c.},
\end{equation}
with $R$ and $S$,
which have mass dimension,
much smaller than the mass $M$ in the $\bar L$-invariant
mass term of Eq.~(\ref{LM}),
but not necessarily smaller than the electroweak scale $v$.
This assumption is natural in the sense of 't Hooft \cite{hooft}.

We have a $5 \times 5$ Majorana mass matrix
following the five chiral neutrino fields in our model.
The neutrino mass term is given by
\begin{equation} \label{mass term}
{\textstyle \frac{1}{2}}\,
\left( \begin{array}{ccccc} \nu_{eL}^T, & \nu_{\mu L}^T, & \nu_{\tau L}^T, &
- \bar\nu_{R1} C, & - \bar\nu_{R2} C \end{array} \right)
\left( \begin{array}{ccccc}
0 & 0 & 0 & a & 0 \\
0 & 0 & 0 & 0 & b^\prime \\
0 & 0 & 0 & 0 & b^{\prime \prime} \\
a & 0 & 0 & R & M \\
0 & b^\prime & b^{\prime \prime} & M & S
\end{array} \right) C^{-1}
\left( \begin{array}{c} \nu_{eL} \\ \nu_{\mu L} \\ \nu_{\tau L} \\
C \bar\nu_{R1}^T \\ C \bar\nu_{R2}^T
\end{array} \right) + {\rm H.c.}
\end{equation}
The mass matrix in Eq.~(\ref{mass term}) has a zero eigenvalue,
corresponding to the normalized eigenvector
\begin{equation} \label{zero eigenvector}
\frac{1}{b} \left( \begin{array}{ccccc} 0, & b^{\prime \prime}, &
- b^\prime, & 0, & 0 \end{array} \right)^T,
\end{equation}
where $b \equiv \sqrt{ \left| b^\prime \right|^2
+ \left| b^{\prime \prime} \right|^2}$.
We thus have two predictions of our model:
there is a massless neutrino ($m_3 = 0$)
and its component along the $\nu_e$ direction vanishes ($U_{e3} = 0$).

We shall start from the fundamental seesaw--soft breaking assumption that
$M$ is much larger than $R$,
$S$,
$a$,
and $b$.
The Majorana mass matrix of Eq.~(\ref{mass term}) then leads to,
besides a zero mass,
two large masses $|\Lambda_\pm|$
and two seesaw-suppressed masses $|\lambda_\pm|$.
We shall, moreover, assume
that $\arg \left( R^\ast S^\ast M^2 \right) = 0$.
This assumption corresponds to the elimination of {\it CP} violation.
Following it we may,
without loss of generality,
use phase transformations of the neutrino fields
in Eq.~(\ref{mass term}) to set $a$,
$b^\prime$,
$b^{\prime \prime}$,
$R$,
$S$,
and $M$ to be real and positive.
This allows us to diagonalize the mass matrix
directly by means of an orthogonal transformation,
\begin{equation} \label{diagonalization}
O^T 
\left( \begin{array}{ccccc}
0 & 0 & 0 & a & 0 \\
0 & 0 & 0 & 0 & b^\prime \\
0 & 0 & 0 & 0 & b^{\prime \prime} \\
a & 0 & 0 & R & M \\
0 & b^\prime & b^{\prime \prime} & M & S
\end{array} \right) O = {\rm diag} \left( \begin{array}{ccccc}
\lambda_+, & \lambda_-, & 0, & \Lambda_+, & \Lambda_- \end{array} \right),
\end{equation}
where the matrix $O$ is orthogonal;
the vector in Eq.~(\ref{zero eigenvector}) constitutes the third column of $O$.
Of course,
the assumption of
{\it CP} conservation has been made only
for the sake of simplicity in the analysis.
Following Eqs.~(\ref{mass term}) and (\ref{diagonalization}), we may write
\begin{equation} \label{relation}
\left( \begin{array}{c} \nu_{eL} \\ \nu_{\mu L} \\ \nu_{\tau L} \\
C \bar\nu_{R1}^T \\ C \bar\nu_{R2}^T \end{array} \right) = O
\left( \begin{array}{c} \nu_{1L} \\ i \nu_{2L} \\ \nu_{3L} \\
\nu_{4L} \\ i \nu_{5L} \end{array} \right) .
\end{equation}
The neutrinos $\nu_1$ and $\nu_2$ are light,
while $\nu_4$ and $\nu_5$ are heavy.
We have inserted factors $i$
in the definition of the fields $\nu_{2L}$ and $\nu_{5L}$; in this way,
these fields are mass eigenfields, because $\lambda_-$ and
$\Lambda_-$ are negative eigenvalues
of the real mass matrix in Eq.~(\ref{diagonalization}). 
Indeed,
$\Lambda_\pm$ and $\lambda_\pm$
may be expanded as series in $M$:
\begin{eqnarray}
\Lambda_\pm &=& \pm M + \frac{R+S}{2} \pm
\frac{4 \left( a^2 + b^2 \right) + \left( R - S \right)^2}{8 M}
+ \ldots,
\label{Lambda} \\
\lambda_\pm &=&
\pm \frac{a b}{M} + \frac{R b^2 + S a^2}{2 M^2} \pm
\frac{R^2 b^4 + S^2 a^4 + 6 R S a^2 b^2 - 4 a^2 b^2 \left( a^2 + b^2 \right)
}{8 a b M^3} + \ldots \label{lambda}
\end{eqnarray}
The normalized eigenvectors
corresponding to the small eigenvalues $\lambda_\pm$ are
\begin{equation} \label{small eigenvectors}
\frac{1}{\sqrt{2}} \left( \begin{array}{c}
{\displaystyle 1 \mp \frac{R b^2 - S a^2}{4 a b M} -
\frac{\left( R b^2 - S a^2 \right)^2 + 16 a^4 b^2}{32 a^2 b^2 M^2}} \\*[3mm]
{\displaystyle \frac{b^\prime}{b} \left[
\mp 1 - \frac{R b^2 - S a^2}{4 a b M}
\pm \frac{\left( R b^2 - S a^2 \right)^2 + 16 a^2 b^4}{32 a^2 b^2 M^2}
\right]} \\*[3mm]
{\displaystyle \frac{b^{\prime \prime}}{b} \left[
\mp 1 - \frac{R b^2 - S a^2}{4 a b M}
\pm \frac{\left( R b^2 - S a^2 \right)^2 + 16 a^2 b^4}{32 a^2 b^2 M^2}
\right]} \\*[3mm]
{\displaystyle \frac{b}{M} \left( \pm 1 + \frac{R b^2 + 3 S a^2}{4 a b M}
\pm \frac{7 S^2 a^4 + 26 R S a^2 b^2 - R^2 b^4 - 16 a^2 b^4 - 32 a^4 b^2}
{32 a^2 b^2 M^2} \right)} \\*[3mm]
{\displaystyle - \frac{a}{M} \left( 1 \pm \frac{3 R b^2 + S a^2}{4 a b M}
+ \frac{7 R^2 b^4 + 26 R S a^2 b^2 - S^2 a^4 - 16 a^4 b^2 - 32 a^2 b^4}
{32 a^2 b^2 M^2} \right)}
\end{array} \right),
\end{equation}
respectively,
and they constitute the first and second columns,
respectively,
of the matrix $O$.
In Eq.~(\ref{small eigenvectors})
we have written down the entries up to sub-subleading order in $M$,
just as in Eq.~(\ref{lambda}).

The mixing angle $\theta_\odot$
relevant for solar-neutrino oscillations is given by
\begin{equation} \label{theta_odot}
\sin^2{2 \theta_\odot}
= \frac{4 \left(  O_{11}  O_{12} \right)^2}
{\left( O_{11}^2 + O_{12}^2 \right)^2}
= 1 - \frac{\left( R b^2 - S a^2 \right)^2}{4 a^2 b^2 M^2}
+ O(M^{-4}).
\end{equation}

The ratio of the mass-squared difference
relevant for solar-neutrino oscillations
over the one relevant for atmospheric neutrinos is,
since $m_3 = 0$,
\begin{equation} \label{ratio}
\frac{\Delta m^2_\odot}{\Delta m^2_{\rm atm}}
= \frac{2 ( \lambda_+^2 - \lambda_-^2 )}{\lambda_+^2 + \lambda_-^2}
= \frac{2 \left( R b^2 + S a^2 \right)}{a b M}
+ O(M^{-2}).
\end{equation}
One sees that the deviation of $\sin^2{2 \theta_\odot}$ from $1$
is quadratic in $\Delta m^2_\odot / \Delta m^2_{\rm atm}$ \cite{akhmedov}.
As the latter quantity is in any case very small---see
Eq.~(\ref{hierarchy})---one concludes that
$\sin^2{2 \theta_\odot}$ is,
for all practical purposes,
$1$.
Thus,
only the LOW and the ``just so'' explanations of the solar-neutrino deficit
may be realized in the context of 
the present model \cite{LOW}.\footnote{The latter explanation is
somewhat
disfavored by the latest results of Super-Kamiokande
\cite{vagins}.}

\vspace*{-2mm}

\paragraph{Orders of magnitude} We have not made any assumption
concerning the relative orders of magnitude of $R$,
$S$,
$a$,
and $b$.
The masses $R$ and $S$ break $\bar L$ softly by two units,
and it is reasonable
to expect $R$ to be of the same order of magnitude as $S$,
although this is by no means required.
On the other hand,
$a$ and $b$ occur in Dirac mass terms,
and it would be reasonable to expect them
to be related to the charged-lepton masses;
as,
however,
$m_e$ and $m_\tau$ differ by three orders of magnitude,
this is not a very useful guideline.
Even if we assume $a = b$ and $R = S$,
we still have some arbitrariness,
since there is a third mass,
namely $M$,
and, of course, the two experimental mass-squared differences
$\Delta m^2_{\rm atm}$ and $\Delta m^2_\odot$
cannot fix the three mass scales in the model.
In any case, Eq.~(\ref{ratio}) tells us that $R/M \sim 10^{-5}$,
if we choose the LOW solution for the solar-neutrino deficit.
Furthermore, with the sensible choice 
$a = b \sim m_\mu \approx 100\ {\rm MeV}$,
one obtains $M \sim 10^8\, {\rm GeV}$
and $R = S \sim 10^3\, {\rm GeV}$.
This result has,
however,
no more than an indicative value
as to what might be the orders of magnitude of $M$,
$R$,
$S$,
$a$,
and $b$.

\vspace*{-2mm}

\paragraph{Bimaximal mixing} 
In our model,
$\sin^2{2 \theta_\odot}$ is almost maximal
as a consequence of the almost exact $\bar L$ symmetry.
In order to reproduce the {\it Ansatz\/}
known as bimaximal mixing \cite{bimaximal},
we would like the $\nu_\mu$--$\nu_\tau$ mixing,
responsible for the depletion of atmospheric muon neutrinos \cite{SK-atm},
to be maximal too.
This would correspond to the massless neutrino,
$\nu_3$,
having equal components along the $\nu_\mu$ and $\nu_\tau$ directions,
i.e.,
to $\left| b^\prime \right| = \left| b^{\prime \prime} \right|$.
Now,
although it is reasonable to expect
$b^\prime$ and $b^{\prime \prime}$ to be of the same order of magnitude,
forcing them to be equal by means of some symmetry is not trivial.
This is because such a symmetry should interchange
$\left( \nu_{\mu L}, \mu_L \right)$
and $\left( \nu_{\tau L}, \tau_L \right)$,
but then it must be broken in order to obtain $m_\mu \neq m_\tau$.
We have been unable to find a way of breaking the interchange symmetry,
and thereby obtaining $m_\mu \neq m_\tau$,
while simultaneously keeping
$\left| b^\prime \right| = \left| b^{\prime \prime} \right|$.
Thus,
it seems impossible to achieve bimaximal mixing {\em naturally\/}
in the context of our model.

\vspace*{-2mm}

\paragraph{Spontaneous breaking of $\bar L$}
Instead of breaking $\bar L$ softly
we may prefer to have it spontaneously broken
through the VEV
of a complex gauge-group singlet $\eta$ with $\bar L = 2$.
Denoting that VEV by $V$,
the Yukawa couplings of $\eta$ are
\begin{equation} \label{eta}
- {\textstyle \frac{1}{2}} \left[
R \left( \eta / V \right) \bar\nu_{R1} C \bar\nu_{R1}^T +
S \left( \eta / V \right)^\ast
\bar\nu_{R2} C \bar\nu_{R2}^T \right] + {\rm H.c.},
\end{equation}
cf.~Eq.~(\ref{soft breaking}).
The spontaneous breaking of the global symmetry $\bar L$
leads to a Goldstone boson,
usually called Majoron.
That massless field is $\sqrt{2}\, {\rm Im}\, \eta$,
if we assume $V$ to be real.
The Majoron is undetectable,
as it couples very weakly to matter \cite{majoron}.
In particular,
its couplings to the light neutrinos are suppressed
by factors $b^2 / M^2$ or $a^2 / M^2$,
while the couplings to the charged leptons and to the quarks
only arise at one-loop order.

\vspace*{4mm}

In conclusion,
we have put forward a simple and economical
extension of the Standard Model which partially explains
the observed features of the neutrino mass spectrum and of lepton mixing.
Our model is based on a $\bar L = L_e - L_\mu - L_\tau$ global symmetry
and on the seesaw mechanism.
The approximate symmetry $\bar L$ is softly broken
by the Majorana mass terms of $\bar L$-charged right-handed neutrinos,
yet all the good predictions of the $\bar L$-symmetric situation
are kept provided the soft-breaking masses are relatively small.
Our model is able to sustain the LOW and the ``just so'' explanations
of the solar-neutrino deficit,
since it predicts maximal solar-neutrino mixing.

\end{document}